\documentclass[conference]{IEEEtran}
\IEEEoverridecommandlockouts
\usepackage{cite}
\usepackage{amsmath,amssymb,amsfonts,extpfeil}
\usepackage{algorithmic}
\usepackage{graphicx}
\usepackage{textcomp}
\usepackage{xcolor}
\def\BibTeX{{\rm B\kern-.05em{\sc i\kern-.025em b}\kern-.08em
    T\kern-.1667em\lower.7ex\hbox{E}\kern-.125emX}}
\usepackage{algorithmic}
\usepackage{algorithm,geometry}
\usepackage{subfigure}
\makeatletter

\newcommand{\Rmnum}[1]{\expandafter\@slowromancap\romannumeral #1@}
\makeatother

\newcommand{\mv}[1]{\mbox{\boldmath{$ #1 $}}}

\begin{document}

\title{Flexible Beam Coverage Optimization for Movable-Antenna Array\\
\author{\IEEEauthorblockN{Dong Wang, Weidong Mei, Boyu Ning, and Zhi Chen}
\IEEEauthorblockA{National Key Laboratory of Wireless Communications \\
University of Electronic Science and Technology of China (UESTC), Chengdu 611731, China
 \\
Emails: DongwangUESTC@outlook.com; wmei@uestc.edu.cn; boydning@outlook.com; chenzhi@uestc.edu.cn}
\thanks{This work was supported by the National Natural Science Foundation of China under Grant 62027806.}
}}

\maketitle

\begin{abstract}
Fluid antennas (FAs) and movable antennas (MAs) have attracted increasing attention in wireless communications recently. As compared to the conventional fixed-position antennas (FPAs), their geometry can be dynamically reconfigured, such that more flexible beamforming can be achieved for signal coverage and/or interference nulling. In this paper, we investigate the use of MAs to achieve uniform coverage for multiple regions with arbitrary number and width in the spatial domain. In particular, we aim to jointly optimize the MAs' weights and positions within a linear array to maximize the minimum beam gain over the desired spatial regions. However, the resulting problem is non-convex and difficult to be optimally solved. To tackle this difficulty, we propose an alternating optimization (AO) algorithm to obtain a high-quality suboptimal solution, where the MAs' weights and positions are alternately optimized by applying successive convex approximation (SCA) technique. Numerical results show that our proposed MA-based beam coverage scheme can achieve much better performance than conventional FPAs.
\end{abstract}

\begingroup
\allowdisplaybreaks
\section{Introduction}
	Beamforming is a pivotal signal-processing technology for multiple-antenna systems, which is able to enhance/reduce the signal strength over desired/undesired directions by optimizing the amplitude and/or phase of the antenna weight vector (AWV)\cite{b1}.  However, due to the fixed geometry of conventional fixed-position antenna (FPA) arrays, their steering vectors have inherent correlation/orthogonality among different steering angles, thus resulting in difficulty in achieving flexible beamforming for interference nulling and/or beam coverage. 
	
	Recently, fluid antennas (FAs) and movable antennas (MAs) have emerged as a new technology that enables local antenna movement within a confined region at the transmitter/receiver, thereby providing additional spatial degrees of freedom (DoFs) to enhance the wireless communication performance\cite{b2,b3,b31,b4,b5}. Inspired by its compelling benefits, there have been some prior studies focusing on its use for point-to-point communications, multi-user systems, secure communicationes, among others\cite{b7,b8,b9,b901,b10,b11,b12}. In particular, the authors in \cite{b13} and \cite{b14} have shown the capability of FAs/MAs to achieve flexible interference nulling, which is able to amplify signals over desired directions and mitigate interference over undesired directions at the same time.
			
	Different from the above papers, we investigate in this paper the use of MAs for flexible beam coverage, aiming to achieve uniform beam gain over a single or multiple regions with different width in the spatial domain, as shown in Fig. \ref{SysMod}. Note that such a beam coverage problem can find broad applications in wireless communications. For example, due to the mobility of users, the beamwidth can be dynamically adjusted to cover an area based on the distribution of the users within it\cite{ning}. Additionally, the number of covered areas may fluctuate over time due to varying network traffic demands. Consequently, more efficient beam coverage approaches are required compared to beamforming design with conventional FPAs only to accommodate diverse requirements for beamwidth and beam number, which motivates this work.
	In particular, we aim to jointly optimize the MAs' weights and positions within a linear array to maximize the minimum beam gain over the desired spatial regions. However, the resulting problem is non-convex and difficult to be optimally solved. To tackle this challenging problem, we propose an alternating optimization (AO) algorithm to obtain a high-quality suboptimal solution to it, where the MAs' weights and positions are alternately updated by applying the successive convex approximation (SCA) technique. Numerical results show that our proposed MA-based beam coverage scheme can achieve much better coverage performance than conventional FPAs.
\begin{figure}
		\centering
		\includegraphics[scale=0.41]{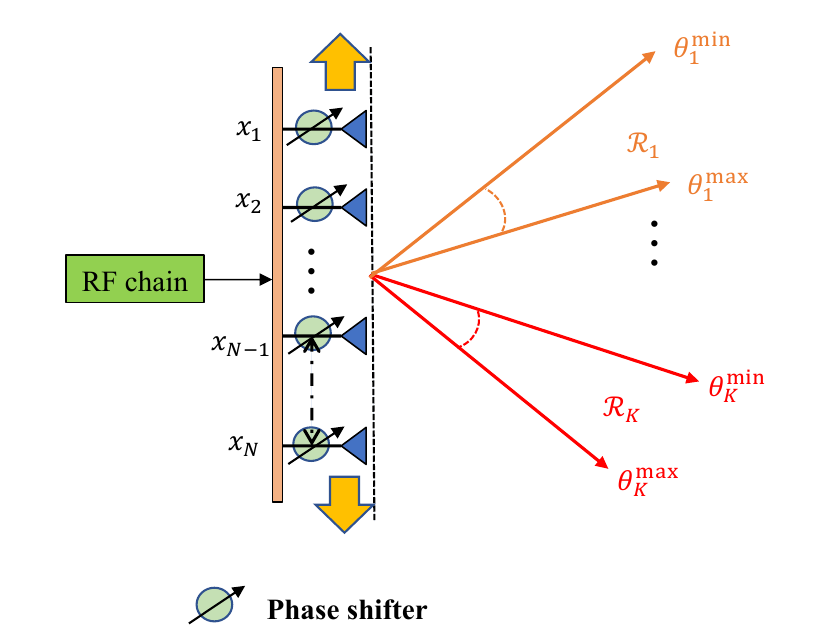}
		\vspace{-6pt}
		\caption{MA-assisted beam coverage for multiple subregions}\label{SysMod}	
		\vspace{-9pt}
\end{figure}	
	    
 {\it Notations:} $\mathbb{R}$ and $\mathbb{C}$ represent the sets of real and complex number, respectively. For a complex number $a$, $\text{Re}\{a\}$ and $\vert a\vert$ denote its real part and modulus, respectively. For a complex-valued vector $\mv{x}$, $\mv{x}^T$, $\mv{x}^H$, $\Vert\mv{x}\Vert_2$ and $\mv{x}(n)$ denote its transpose, conjugate transpose, $l_2$-norm, $n$-th element, respectively. For a matrix $\mv{A}$, $\text{Tr}(\mv{A})$ represents its trace. $\mv{x}\sim\mathcal{CN}(\mv{\mu},\mv{R})$ represents that $\mv{x}$ follows circularly symmetric complex Gaussian (CSCG) distribution with mean $\mu$ and covariance matrix $\mv{R}$.

\section{System Model}
As depicted in Fig. 1, we consider a transmitter equipped with a linear array with $N$ MAs, whose positions can be flexibly adjusted within a one-dimensional (1D) region denoted as $\mathcal{C}_t$. The length of the MA array is set to $D$. For convenience, we assume a reference origin point $x_0=0$ in the transmit region $\mathcal{C}_t$, as shown in Fig. 1. Let $x_n$ denote the coordinate of the $n$-th MA's position, $n\in \mathcal N \triangleq \{1,2,\cdots,N\}$. Accordingly, the antenna position vector (APV) of all $N$ MAs is denoted by $\mv{x}\triangleq[x_1,x_2,\cdots,x_N]\in \mathbb{R}^{N\times1}$. Let $f_c$ and $\lambda$ denote the operating carrier frequency of the considered system and carrier wavelength, respectively. For a given steering angle $\theta$, the array response vector of the MA array can be determined as a function of the APV $\mv{x}$ and $\theta$, i.e.,
\begin{align}
    \mv{a}(\mv{x},\theta) = [e^{j\frac{2\pi}{\lambda}x_1\cos\theta},\cdots,e^{j\frac{2\pi}{\lambda}x_N\cos\theta}]^T\in \mathbb{C}^{N\times 1}. 
\end{align}

To ease the practical implementation, we assume analog beamforming with a constant magnitude at the transmitter, which is given by
\begin{equation}
    \mv{\omega} = \frac{1}{\sqrt{N}}[e^{j\phi_1},e^{j\phi_2},\cdots,e^{j\phi_N}]^T\in \mathbb{C}^{N\times1}, 
\end{equation}
where $\phi_n$ denotes the phase shift at the $n$-th MA.
 
Based on the above, the beam gain over the steering angle $\theta$ can be expressed as
\begin{equation}
    G(\mv{\omega},\mv{x},\theta)=\big|\mv{\omega}^H\mv{a}(\mv{x},\mathcal{\theta})\big|^2.
\end{equation}
	
As shown in Fig. 1, in this paper, we aim to achieve uniform beam coverage in the angular domain by jointly optimizing the APV $\mv{x}$ and the analog beamformer $\mv{\omega}$. In particular, we consider that the desired coverage region is composed of $K$ disjoint subregions, i.e.,  $\mathcal{R}=\mathcal{R}_1\cup\mathcal{R}_2\cup\cdots\cup\mathcal{R}_K$, with $\mathcal{R}_k=\{\theta\in[\theta_{\min}^{(k)},\theta_{\max}^{(k)}]\}$, where $\theta_{\min}^{(k)}$ and $\theta_{\max}^{(k)}$ denote the boundary angles of the $k$-th subregion, with $\theta_{\min}^{(k)} < \theta_{\max}^{(k)}$. It follows that by changing the number of the subregions (i.e., $K$) and the width of each subregion $k$ (i.e., $\theta_{\max}^{(k)} - \theta_{\min}^{(k)}$), flexible beam coverage can be achieved.

To achieve uniform beam coverage over the $K$ subregions, we aim to maximize the minimum beam gain over them, i.e.,
\begin{equation}\label{g_min}
    G_{\min}(\mv{\omega},\mv{x})=\min_{\theta\in\mathcal{R}}G(\mv{\omega},\mv{x},\theta),
\end{equation}
by jointly optimizing the APV $\mv{x}$ and the analog beamforming vector $\mv{\omega}$. The associated optimization problem can be formulated as
\begin{subequations}
\begin{align}
    (\text{P1}) &\quad\max_{\mv{\omega},\mv{x}}G_{\min}(\mv{\omega},\mv{x})\nonumber
    \\
    \mbox{s.t.} &\quad 0\leq x_n\leq D,\quad n\in\mathcal{N}, \label{p1b}
    \\
    &\quad x_n-x_{n-1}\geq d_{\min},\quad n\in\mathcal{N}\backslash\{1\},\label{p1c}
    \\
    & \quad  \vert\mv{\omega}(n)\vert=\frac{1}{\sqrt{N}},\quad n\in \mathcal{N,}\label{p1d}
\end{align}
\end{subequations}
where constraint (\ref{p1b}) ensures that the MAs are located within the transmit region $\mathcal{C}_t$; (\ref{p1c}) ensures the minimum distance between adjacent MAs (denoted as $d_{\min}$) to avoid antenna coupling; constraint (\ref{p1d}) is with respect to (w.r.t.) the non-convex modulus constraints on the analog beamforming vector. 

However, (P1) is generally difficult to be optimally solved due to the continuous coverage region, the coupling of the analog beamformer $\mv{\omega}$ and APV $\mv{x}$ in objective function, as well as the unit-modulus constraints. Moreover, the objective function appears to be non-convex in both $\mv{\omega}$ and $\mv{x}$. Next, we propose an AO algorithm to solve (P1) iteratively.

\section{Proposed algorithm}
First, due to the continuous nature of the angle within each subregion $\mathcal{R}_k$, we first discretize it into $L_k$ discrete values, i.e.,
\begin{align}
	\theta_l^{(k)} = \theta_{\min}^{(k)}+\frac{l^{(k)}-1}{L^{(k)}-1}Z_k,l^{(k)}=1,\cdots,L^{(k)},
\end{align}
where $L^{(k)}$ represent the number of the sampling points in $\mathcal{R}_k$, and $Z_k=\theta_{\max}^{(k)}-\theta_{\min}^{(k)}$ represents the width of $\mathcal{R}_k$.

Next, we introduce an auxiliary variable $t$ to convert (P1) into its epigraph form, i.e.,
\begin{subequations}
\begin{align}
	(\text{P2}) &\quad \max_{\mv{\omega},\mv{x},t} t \nonumber
	\\
	\mbox{s.t.} &\quad 
	G(\mv{\omega},\mv{x},\theta_l^{(k)})\geq t, \forall k,l
	\label{p2b}\\
	& \quad (\text{\ref{p1b}}),(\text{\ref{p1c}}),(\text{\ref{p1d}})\nonumber.
\end{align}
\end{subequations}

However, (P2) is still difficult to solve due to its non-convex constraints w.r.t. both $\mv{w}$ and $\mv{x}$. Next we propose an AO algorithm to optimize them alternately with the SCA technique.

\subsection{Optimization of $\mv{\omega}$ for Given $\mv{x}$}
First, we optimize the analog beamforming vector $\mv{\omega}$ in (P2) with a given APV $\mv{x}$. Then, (P1) can be simplified as
\begin{align}
	(\text{P2.1}) &\quad \max_{\mv{\omega},t} t \nonumber
	\\
	\mbox{s.t.} & \quad (\text{\ref{p2b}}),(\text{\ref{p1d}}). \nonumber
\end{align}
To deal with the non-convex constraint (\ref{p2b}) , we first rewrite the beam gain at $\theta_l^{(k)}$ as 
\begin{align}
	G(\mv{\omega},\mv{x},\theta_l^{(k)})&=\big\lvert\mv{\omega}^H\mv{a}(\mv{x},\theta_{l}^{(k)})\mv{a}^H(\mv{x},\theta_l^{(k)}) \mv{\omega}\big\rvert^2
	\nonumber\\
	&=\text{Tr}(\mv{R}_l^{(k)}\mv{V}),
\end{align}
where $\mv{R}_l^{(k)}=\mv{a}(\mv{x},\theta_{l}^{(k)})\mv{a}^H(\mv{x},\theta_k^{(l)})$ and $\mv{V}=\mv{\omega}\mv{\omega}^H$, with $\text{rank}(\mv{V})=1$. If we apply the semidefinite relaxation (SDR) to relax the rank-one constraint, (P2.1) can be recast as
\begin{subequations}
\begin{align}
	(\text{P2.2})\quad &\max_{t,\mv{V}} t
	\nonumber\\
	\mbox{s.t.} \quad & \text{Tr}(\mv{R}_l^{(k)}\mv{V})\geq t,\quad\forall l, k\label{p211b}
	\\
	&\mv{V}(n,n)=\frac{1}{N}, \quad n\in\{1,\cdots,N\}\label{p211c}.
	\\
	&\mv{V} \succeq \mv{0}\label{p211d},
\end{align}
\end{subequations}
Note that as (\ref{p211b}), (\ref{p211c}) and (\ref{p211d}) are all affine functions, (P2.2) is a classical semi-definite program (SDP) problem, which thus can be optimally solved via the interior-point algorithm\cite{b12}. However, the optimized matrix $\mv{V}$ may not be rank-one. Although Gaussian randomization procedures may be applied to yield a rank-one solution, the resulting performance gap with the globally optimal solution may be large. To tackle this issue, we propose a penalty-based algorithm in this paper, which moves the rank-one constraint to the objective function. Specifically, note that the rank-one constraint can be equivalently written as
\begin{align}
	\mbox{rank}(\mv{V})=1 \Leftrightarrow f(\mv{V})\triangleq\text{Tr}(\mv{V})-\sigma(\mv{V})=0,\label{rank-one}
\end{align}
 where $\sigma({\mv{V}})$ represents the maximum singular value of $\mv{V}$. As such, we can consider the following objective function by adding a penalty term, i.e.,
\begin{align}
	\max_{t,\mv{V}}t-\rho f(\mv{V})\label{pen_obj},
\end{align}
where $\rho\geq 0$ represents a penalty parameter ensuring that the objective function is small enough if $\text{Tr}(\mv{V})-\sigma_{\max}\neq 0$. However, such an objective function is still non-convex as the maximum singular value of $\mv{V}$, $\sigma(\mv{V})$, is a convex function (instead of being concave) in $\mv{V}$. Nonetheless, it enables us to apply the SCA algorithm to obtain a locally optimal solution.  In particular, 
	for a given local point $\mv{V}^{(i)}$ in the $i$-th SCA iteration, we replace $f(\mv{V})$ as its first-order Taylor expansion, i.e., 
	\begin{align}
		f(\mv{V})\geq \tilde f(\mv{V}|\mv{V}^{(i)})\triangleq\text{Tr}&(\mv{V})-\sigma(\mv{V}^{(i)})+\nonumber\\&\text{Re}\{\text{Tr}(\mv{s}\mv{s}^H)(\mv{V}-\mv{V}^{(i)})\},\label{obj_new}
	\end{align}
where $\mv{s}$ represents the singular vector corresponding to the maximum singular value of $\mv{V}^{(i)}$. where the equality holds only when $\mv{V}=\mv{V}^{(i)}$. Based on the above, the optimization of $\mv{V}$ in the $i$-th SCA iteration is given by
\begin{align}
	(\text{P2.3})\quad &\max_{t,\mv{V}}  t-\rho \tilde f(\mv{V}|\mv{V}^{(i)})
	\nonumber\\
	& \mbox{s.t.} \quad(\text{\ref{p211b}}),(\text{\ref{p211c}}),(\text{\ref{p211d}}).\nonumber
\end{align}
It can be seen that the objective function of (P2.3) is currently a linear function of $t$ and $\mv{V}$. As such, (P2.3) is an SDP problem which can be optimally solved\cite{b16}. Based on our simulation results, by properly setting the value of $\rho$, the converged solution of $\mv{V}$ is always rank-one. Next, we proceed to the $(i+1)$-th SCA for $\mv{V}$ by updating $\mv{V}^{(i+1)}=\mv{V}^{(i)}$. Finally, to extract $\mv{\omega}$ based on the converged $\mv{V}$, we can perform the singular value decomposition (SVD) on $\mv{V}$ as $\mv{V}=\mv{U}_1^H\mv{\Lambda}\mv{U}_2$, where $\mv{U}_1$, $\mv{\Lambda}$ and $\mv{U}_2$ are the left eigenvector matrix, the diagonal matrix of the singular values and the right eigenvector matrix of $\mv{V}$, respectively. Then we can obtain the analog beamforming solution to (P2.1) as $\mv{\omega}=1/\sqrt{N}e^{j\arg({\mv{U}_1^H\mv{\Lambda}^{1/2}\mv{r}})}$, where $\mv{r}$ is the left singular vector corresponding to the largest singular value of $\mv{V}$.

\subsection{Optimization of $\mv{x}$ for Given $\mv{\omega}$}
Next, we optimize the APV $\mv{x}$ in (P2) with a given analog beamformer 
  $\mv{\omega}$. To tackle the non-convex constraint (\ref{p2b}), we first expand $G(\mv{\omega},\mv{x},\theta_l^{(k)})$ into the following form
  \begin{align}
  	G&(\mv{\omega},\mv{x},\theta_{l}^{(k)})=\big\vert\mv{\omega}^H\mv{a}(\mv{x},\theta_{l}^{(k)})\big\vert^2\nonumber\\
  	&=\sum_{p=1}^{N}\sum_{q=1}^{N}\frac{1}{N}\cos(\frac{2\pi}{\lambda}\cos(\theta_{l}^{(k)})(x_p-x_q)-(\phi_p-\phi_q))\nonumber\\
  	&\triangleq\sum_{p=1}^{N}\sum_{q=1}^{N}\frac{1}{N}\cos(u_l^{(k)}(x_p,x_q)).\label{cos} 
  \end{align}
where $\cos(u_l^{(k)}(x_p,x_q))\triangleq (\alpha_l^{(k)})^2(x_p-x_q)-(\phi_p-\phi_q)$ and $\alpha_l^{(k)}=\frac{2\pi}{\lambda}\cos(\theta_l^{(k)})$. The difficulty in dealing with (14) lies in the cosine function therein. To tackle this challenge, we relax (\ref{cos}) by using the SCA again. For a given $z_0\in\mathbb{R}$, its second-order Taylor expansion can be expressed as
\begin{align}
	\cos(z_0)-\sin(z_0)(z-z_0)-\frac{1}{2}\cos(z_0)(z-z_0)^2.
\end{align}
As $\cos(z_0)\leq1$ and $(z-z_0)^2\geq0$, we can construct a quadratic function $g(z|z_0)$ to replace $\cos(z)$ as
\begin{align}
	\cos(z)\geq g(z|z_0)\triangleq\cos(z_0)-\sin(z_0)(z-z_0)-\frac{1}{2}(z-z_0)^2.\label{gz}
\end{align}
Based on the above, for any given $\mv{x}^{(i)}$ in the $i$th iteration of SCA for the APV $\mv{x}$, by substituting (\ref{gz}) into (\ref{cos}), we can construct a surrogate function of $G(\mv{\omega},\mv{x},\theta_{k})$ as
\begin{align}
	G(\mv{\omega},\mv{x},\theta_l^{(k)})&\geq\sum_{p=1}^N\sum_{q=1}^{N}\frac{1}{N}g(u_l^{(k)}(x_p,x_q)|u_l^{(k)}(x_p^{(i)},x_q^{(i)}))\nonumber\\
	&\triangleq \mv{x}^T\mv{A}_l^{(k)}\mv{x}+[\mv{b}_l^{(k)}]^T\mv{x}+c_l^{(k)},
\end{align}
where $\mv{A}_l^{(k)}\in\mathbb{R}^{N\times N}$, $\mv{b}_l^{(k)}\in\mathbb{R}^{N\times 1}$, and $c_l^{(k)}\in\mathbb{R}$ are given by
\begin{equation}
	\mv{A}_l^{(k)}=-(\alpha_l^{(k)})^2(\mv{I}_N-\frac{1}{N}\mv{1}_N)\triangleq-(\alpha_l^{(k)})^2\mv{W},\label{A}
\end{equation}
\begin{align}
		\mv{b}_l^{(k)}(p)=\frac{2\alpha_l^{(k)}}{N}\sum_{q=1}^{N}[\alpha_l^{(k)}x_{p,q}^{(i)}-\sin(u_l^{(k)}(x_p^{(i)},x_q^{(i)}))],\label{b}
\end{align}
\begin{align}
		c_l^{(k)}=&\frac{1}{N}\sum_{p=1}^{N}\sum_{q=1}^{N}[\cos(u_l^{(k)}(x_p^{(i)},x_q^{(i)}))+\nonumber\\&\alpha_l^{(k)}x_{p,q}^{(i)}\sin(u_l^{(k)}(x_p^{(i)},x_q^{(i)}))-\frac{1}{2}(\alpha_l^{(k)}x_{p,q}^{(i)})^2],
\end{align}
where $\mv{I}_N$ and $\mv{1}_N$ represent the identity matrix and all-ones matrix of size $N$, respectively, $x_{p,q}^{(i)}\triangleq x_p^{(i)}-x_q^{(i)}$, and $\mv{W}\triangleq\mv{I}-\frac{1}{N}\mv{1}_N$. As such, constraint (\ref{p2b}) can be relaxed as
\begin{align}
	\mv{x}^T\mv{A}_l^{(k)}\mv{x}+[\mv{b}_k^{(l)}]^T\mv{x}+c_l^{(k)}\geq t, \forall k,l.\label{qc}
\end{align}
Next, we show that (21) is a quadratic constraint (QC). To this end, it is noted that both $\mv{A}_l^{(k)}$ and $\mv{W}$ are hermitian matrices for any given $k$ and $l$ and have the same eigenvectors. Thus, $\mv{A}_l^{(k)}$ can be diagonalized as
\begin{align}
	\mv{A}_l^{(k)}=[\mv{U}_l^{(k)}]^H\mv{D}_l^{(k)}\mv{U}_l^{(k)}\triangleq-[\mv{U}_l^{(k)}]^H(\alpha_l^{(k)})^2\mv{D}\mv{U}_l^{(k)}\end{align}
where $\mv{U}_l^{(k)}$ is an unitary matrix composed of the eigenvectors of $\mv{A}_l^{(k)}$, $\mv{D}_l^{(k)}$ and $\mv{D}$ are diagonal matrices composed of the eigenvalues of $\mv{A}_l^{(k)}$ and $\mv{W}$, respectively, with $\mv{D}_l^{(k)}=-(\alpha_l^{(k)})^2\mv{D}$. Furthermore, any eigenvalue of $\mv{D}$ is non-negative according to Gerschgorin disk theorem. Hence, the eigenvalue of $\mv{D}_l^{(k)}$ is non-positive and $\mv{A}_l^{(k)}$ is a negative semi-definite matrix. It follows that constraint (\ref{qc}) is a convex QC w.r.t. $\mv{x}$.

Therefore, in the $i$-th iteration of the SCA for $\mv{x}$, $\mv{x}$ can be optimized by solving the following optimization problem
\begin{subequations}
\begin{align}
	(\text{P2.4}) &\max_{\mv{x},t}t\nonumber\label{p22a}\\
	\mbox{s.t. } & (\text{\ref{p1b}}),(\text{\ref{p1c}}),(\text{\ref{qc}})\nonumber.
\end{align}
\end{subequations}
Since constraints (\ref{p1b}) and (\ref{p1c}) are linear and (\ref{qc}) is a QC w.r.t \mv{x}, (P2.2) is a classic quadratically constrained quadratic program (QCQP) problem which can be effectively solved via the interior-point algorithm. Next, we proceed to the $(i+1)$-th SCA for $\mv{x}$ by updating $\mv{x}^{(i+1)}=\mv{x}^{(i)}$.

\subsection{Overall Algorithm and Complexity Analysis}\label{analysis}
The overall AO algorithm can be executed as follows. Let 
	$\mv{\omega}{(j-1)}$ and $\mv{x}{(j-1)}$ denote the values of $\mv{\omega}$ and $\mv{x}$ at the beginning of the $j$-th AO iteration. Then, in this AO iteration, we first optimize $\mv{V}$ via SCA with fixing $\mv{x}=\mv{x}(j-1)$ and $\mv{V}^{(0)}=\mv{\omega}(j-1)\mv{\omega}^H (j-1)$ and obtain $\mv{\omega}(j)$ by performing SVD on the converged $\mv{V}$. 
	Next, we optimize $\mv{x}$ via SCA with fixing $\mv{\omega}=\mv{\omega}(j)$ and $\mv{x}^{(0)}=\mv{x}(j-1)$ and obtain $\mv{x}(j)$ as the converged solution, and the $(j+1)$-th AO iteration follows. The overall procedures of the AO algorithm are summarized in Algorithm 1.
\begin{algorithm}[!h]
    \caption{Proposed AO algorithm to solve (P1)}
    \label{alg:AOA}
    \renewcommand{\algorithmicrequire}{\textbf{Input:}}
    \renewcommand{\algorithmicensure}{\textbf{Output:}}
    \begin{algorithmic}[1]
        \REQUIRE  $\mv{\omega}(0)$ and $\mv{x}(0)$. 
        \ENSURE $\mv{\omega}$, $\mv{x}$.    
        \STATE Initialization: $j\leftarrow 1$.
        \WHILE{AO convergence is not reached}
		\STATE Initialize $i\leftarrow 0$ and update $\mv{V}^{(0)}=\mv{\omega}(j-1)\mv{\omega}^H(j-1)$, and $\mv{x}=\mv{x}(j-1)$.
		\WHILE{SCA convergence for $\mv{V}$ is not reached}	
		\STATE Obtain $\mv{V}^{(i+1)}$ by solving problem (P2.3).
		\STATE Update $i\leftarrow i+1$.		
		\ENDWHILE
		\STATE Obtain $\mv{\omega}(j)$ based on the SVD on $\mv{V}^{(i)}$.		
		\STATE Initialize $i\leftarrow 0$ and update $\mv{\omega}=\mv{\omega}(j)$, $\mv{x}^{(0)}=\mv{x}(j-1)$.
		\WHILE{SCA convergence for \mv{x} is not reached}
		\STATE Obtain $\mv{x}^{(i+1)}$ by solving problem (P2.4).
		\STATE Update $i\leftarrow i+1$.
		\ENDWHILE
		\STATE Update $\mv{x}(j)=\mv{x}^{(i)}$.
		\STATE $j\leftarrow j+1$.
		\ENDWHILE       
		\RETURN $\mv{x}$ and $\mv{\omega}$.
    \end{algorithmic}
\end{algorithm}

Next, we prove that the proposed AO algorithm with SCA is ensured to converge. Specifically, regarding the SCA for $\mv{V}$, let $v^{(i)}\triangleq t-\rho f(\mv{V}^{(i)})$ denote the objective value of (11) in the $i$-th SCA iteration for $\mv{V}$. Then, the following inequalities hold, i.e.,
\begin{align}
	v^{(i)}&\triangleq t-\rho f(\mv{V}^{(i)})\overset{(a)}{=}t-\rho\tilde f(\mv{V}^{(i)}|\mv{V}^{(i)})
	\\ & \overset{(b)}{\leq} t-\rho\tilde f(\mv{V}^{(i+1)}|\mv{V}^{(i)})\overset{(c)}{\leq} t-\rho f(\mv{V}^{(i+1)})\triangleq v^{(i+1)},\nonumber
\end{align}
where the equality $(a)$ holds since the first-order Taylor expansion in (\ref{obj_new}) is tight at $\mv{V}^{(i)}$; the inequality $(b)$ holds as $\mv{V}^{(i+1)}$ is the optimal solution to (P2.3) and thus maximizes the function $t-\rho \tilde f(\mv{V}|\mv{V}^{(i)})$; the inequality $(c)$ holds due to (12). Based on the above, the sequence $\{v^{(i)}\}$ is non-decreasing and thus ensured to converge. Similarly, it can be shown that the SCA for optimizing the APV $\mv{x}$ also converges, for which the details are omitted for brevity. It follows that the proposed AO algorithm in Algorithm 1 must converge.

Finally, we analyze the complexity Algorithm 1. It can be shown that the complexity of optimizing the analog beamformer $\mv{\omega}$ and the APV $\mv{x}$ are both in the order of $\mathcal{O}(\sqrt{L}N(N^2+L))$\cite{o}, where $L=\sum_{k=1}^{K}L^{(k)}$ denotes the total number of sampling points.

\subsection{Initialization}\label{initia}
The converged performance of the AO algorithm depends critically on the initialization. In this subsection, we propose an efficient initialization method to detemine $\mv{x}(0)$ and $\mv{\omega}(0)$ in Algorithm 1. For the initial MA positions, we consider that all MAs are uniformly deployed within the linear array with equal spacing. As such, we set $\mv{x}(0)=[\frac{D}{N+1},\frac{2D}{N+1},\cdots,\frac{ND}{N+1}]^T$.

Next, to determine $\mv{\omega}(0)$, we consider solving the SDR problem (P2.2) directly with fixing $\mv{x}=\mv{x}(0)$ via the interior-point algorithm. Denote by $\mv{V}(0)$ the associated optimal solution to (P2.2). Then, we obtain $\mv{\omega}(0)$ by performing Gaussian randomization based on $\mv{V}(0)$, i.e., 

\begin{align}
	\mv{\omega}(0)=\frac{1}{\sqrt{N}}e^{j\arg{(\mv{U}_0^H\mv{\Lambda}_0^{1/2}\mv{r}_0})},
\end{align}
where $\mv{U}_0$ and $\mv{\Lambda}_0$ are unitary matrix and diagonal matrix composed by eigenvectors and eigenvalues of the matrix $\mv{V}(0)$, respectively, and vector $\mv{r}_0\in\mathbb{C}^{N\times 1}$ is a random vector following $\mathcal{CN}(0,\mv{I}_{N})$. By comparing the objective value of (P2.2) under different realizations of $\mv{r}_0$, we set $\mv{\omega}(0)$ as that corresponding to the one yielding the maximum objective value of (P2.2).

\begin{figure*}[hbtp]
\centering
\subfigure[]{\includegraphics[width=0.4\textwidth]{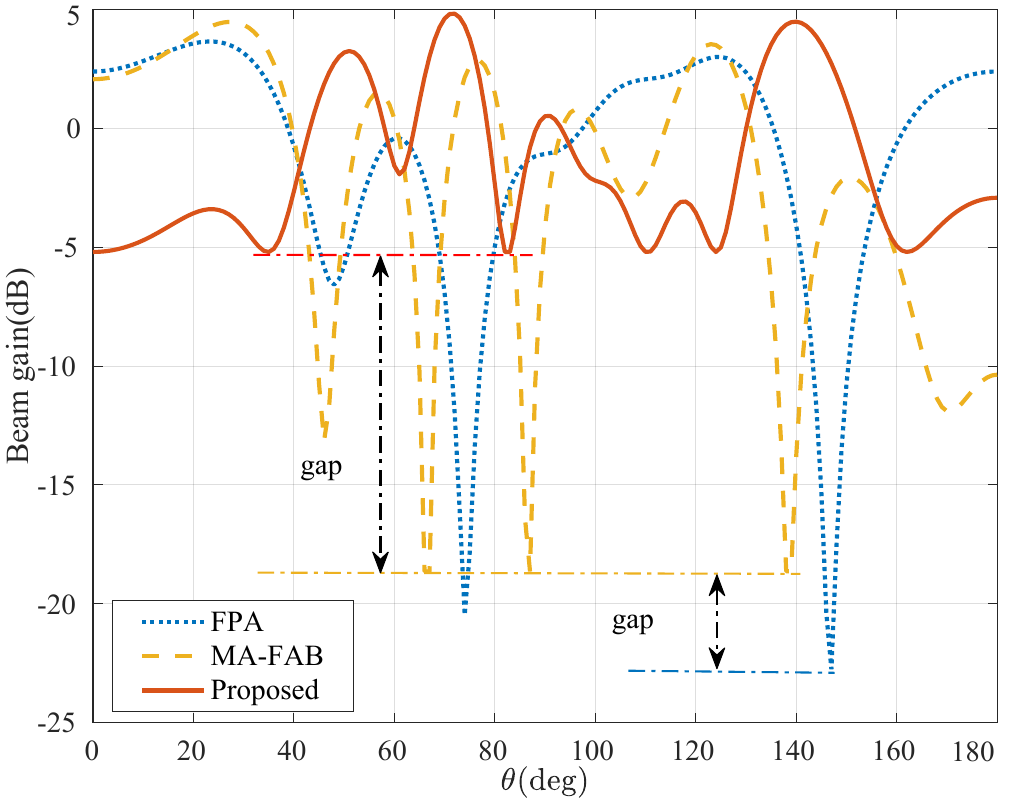}\label{n_6}}
\qquad\quad
\subfigure[]{\includegraphics[width=0.42\textwidth]{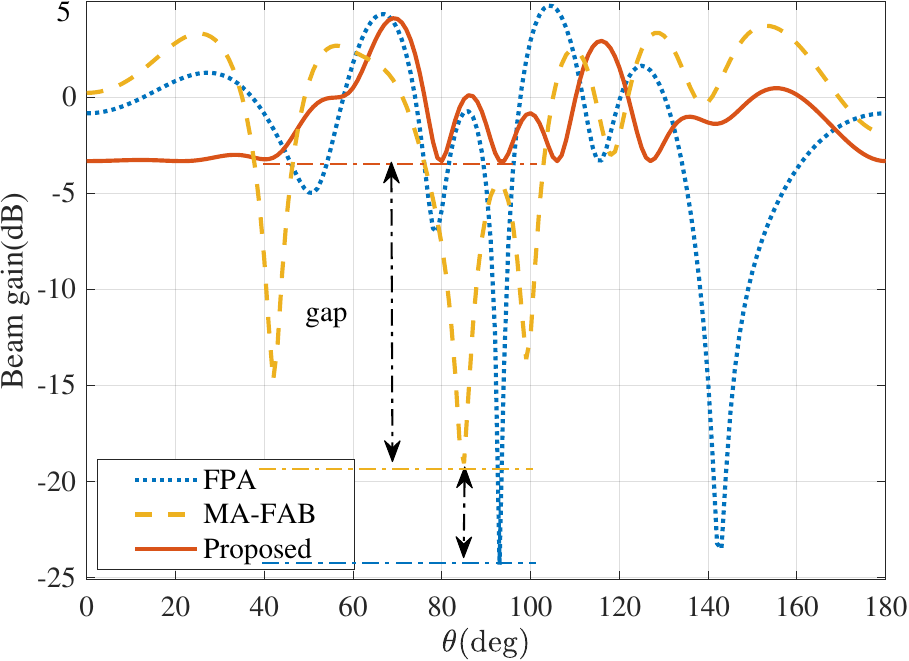}\label{n_8}}
\subfigure[]{\includegraphics[width=0.42\textwidth]{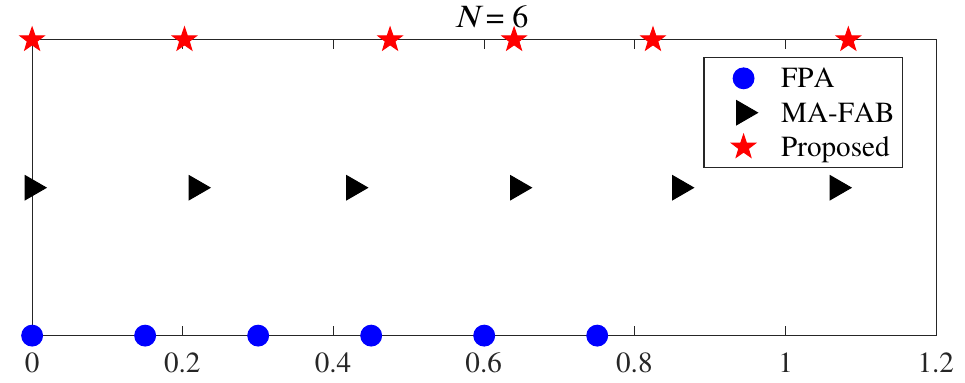}\label{apv1}}
\qquad
\subfigure[]{\includegraphics[width=0.42\textwidth]{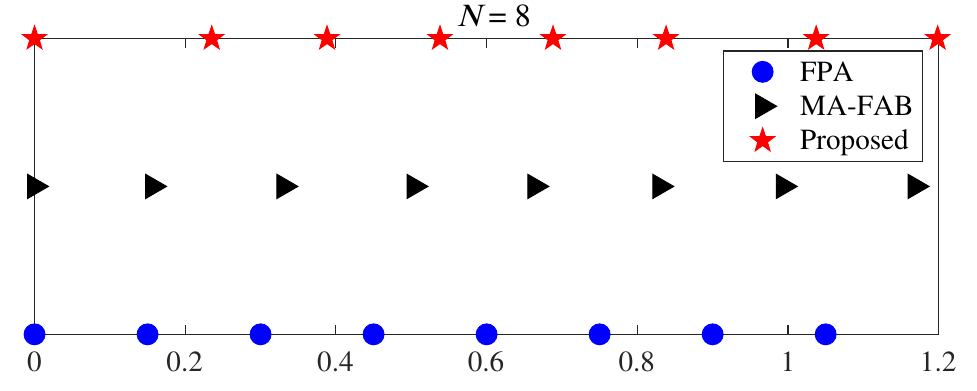}\label{apv2}}
\vspace{-8pt}
\caption{Optimized beam gains with (a) $N=6$ and (b) $N=8$, and optimized MA positions with (c) $N=6$ and (d) $N=8$.}\label{simfig}
\vspace{-12pt}
\end{figure*}

\section{Numerical Results}
In this section, we provide numerical results to evaluate the performance of our proposed MA-enabled beam coverage. Unless otherwise stated, the simulation parameters are set as follows. The penalty parameter for updating AWV is set to $\rho=20$. The carrier frequency is $f_c=1$ GHz. The length of the MA linear array is set to $D=8\lambda$. Meanwhile, we set the convergence threshold for the AO algorithm, the SCA for \mv{V}, and the SCA for \mv{x} as $10^{-5}$, $0.01$, and $0.01$, respectively. For performance comparison, we consider the following two benchmarks:

1) \textbf{FPA}: $N$ FPAs are deployed with half-wavelength spacing and the analog beamfroming is optimized based on the SCA algorithm as in Section III-A.

2) \textbf{MAs with fixed analog beamforming (MA-FAB)}: In this benchmark, we set the analog beamforming $\mv{\omega}=\mv{\omega}(0)$ as presented in Section III-D and optimize MA positions based on the SCA algorithm as in Section III-B.
 
 First, by setting $K=1$, $\theta_{\min}^{(1)}=0$, and $\theta_{\max}^{(1)}=180^{\circ}$, i.e., $\mathcal{R}=[0,180^{\circ}]$, Figs. \ref{n_6} and \ref{n_8} plot the optimized beam gains over $\cal R$ by different schemes for $N=6$ and $N=8$, respectively. Note that we have introduced a common shift to the MA positions such that $x_1=0$ without affecting the coverage performance. The optimized MA positions by different schemes are also shown in Figs. \ref{apv1} and \ref{apv2} for $N=6$ and $N=8$, respectively. It is observed from Figs.\,\ref{n_6} and \ref{n_8} that our proposed AO algorithm can achieve a considerably higher max-min beam gain over $\mathcal{R}$ as compared to the two benchmarks, thereby resulting in flatter beam gain for both $N=6$ and $N=8$. 
 It is also observed that the MA-FAB benchmark outperforms the FPA benchmark, which indicates that antenna positions may play a more significant role than antenna weights in terms of beam coverage. 
 In addition, it is observed from Figs. \ref{apv1} and \ref{apv2} that the optimized MA positions are different in different schemes. In particular, for both $N=6$ and $N=8$, the proposed scheme and the MA-FAB benchmark result in a larger antenna aperture, i.e., $x_N-x_1$, than the FPA benchmark, so as to fully exploit the spatial DoFs available. Moreover, the inter-antenna spacing is observed to be non-uniform in the former two schemes employing MAs, unlike the FPA benchmark with an equal antenna spacing. In addition, by comparing Figs. 2(c) and 2(d), it is observed that the antenna aperture under $M=8$ is larger than that under $M=6$ in the proposed scheme (1.2 versus 1.14), so as to accommodate the movement of more MAs.   
      
Next, Fig.\,\ref{multi_zone} plots the optimized beam gain over $\mathcal{R}$ by setting $K=3$, $\mathcal{R}_1=[0,30^{\circ}]$, $\mathcal{R}_2=[70^{\circ},110^{\circ}]$ and $\mathcal{R}_3=[160^{\circ},170^{\circ}]$. The number of MAs is set to $N=8$. The optimized MA positions by the considered schemes are shown in Fig.\,\ref{apv3}. It is observed that in the case of multi-region coverage, the proposed scheme still achieves much higher max-min beam gain than the other two benchmarks. In particular, the minimum beam gain by our proposed scheme is observed to be even larger than the maximum beam gain by the FPA benchmark in $\mathcal{R}_1$ and $\mathcal{R}_3$. While in $\mathcal{R}_2$, although the FPA can achieve a maximum beam gain of $6$ dB, its minimum beam gain is considerably lower ($-19$ dB) than that by the proposed scheme ($-1$ dB). Moreover, the MA-FAB benchmark is observed to achieve a slightly higher minimum beam gain than the proposed scheme in $\mathcal{R}_2$. This implies that as far as each subregion concerned, the proposed scheme may not achieve the highest minimum beam gain. It is also observed from Fig.\,\ref{apv3} that similar to Figs.\,\ref{apv1} and \ref{apv2}, the proposed scheme and the MA-FAB benchmark result in a larger antenna aperture than the FPA benchmark. Particularly, $x_8$ increases from 1.2 in Fig.\,\ref{apv2} to 1.65 in Fig.\,\ref{apv3}, which suggests that multi-region coverage may demand a larger antenna aperture than its single-region counterpart due to the discontinuous subregions. Interestingly, it is observed that the optimized MA positions by the proposed scheme can be viewed as being composed of three subarrays, with the first and second MAs in the first subarray, the third to fifth MAs in the second subarray, and the sixth to eighth MAs in the third subarray, which helps cover the three subregions.  
\begin{figure}[hbtp]
\centering
\subfigure[]{\includegraphics[width=0.4\textwidth]{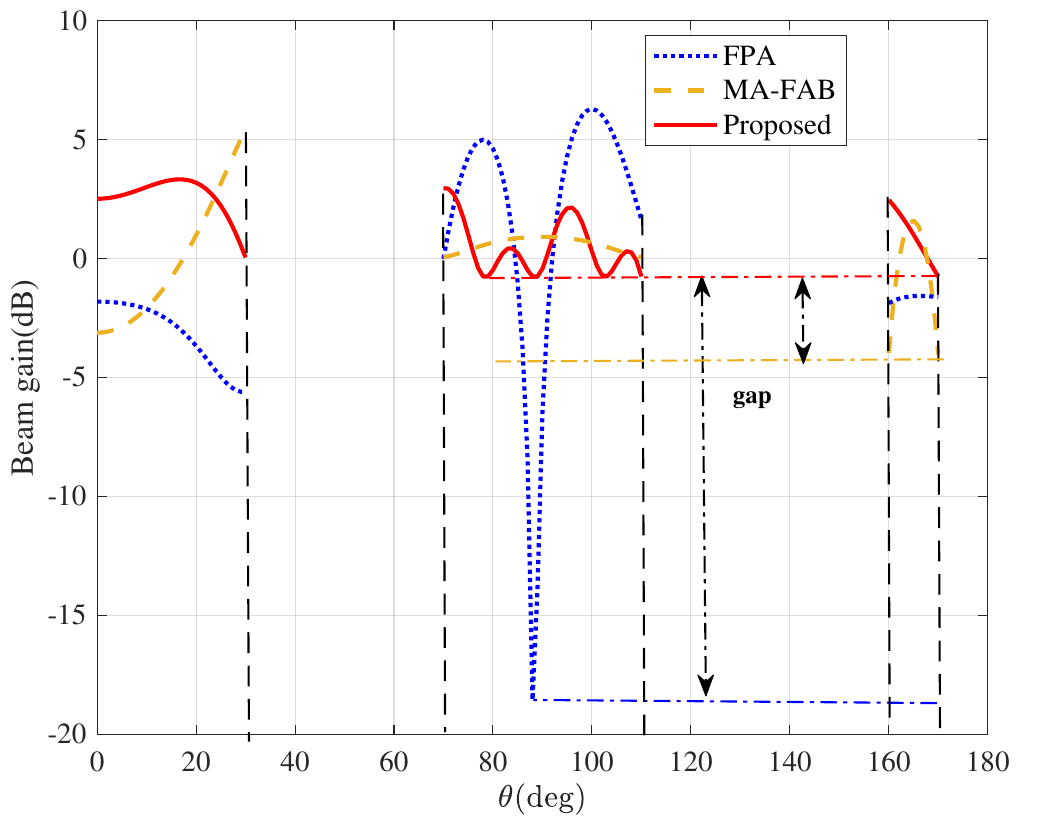}\label{multi_zone}}
\subfigure[]{\includegraphics[width=0.4\textwidth]{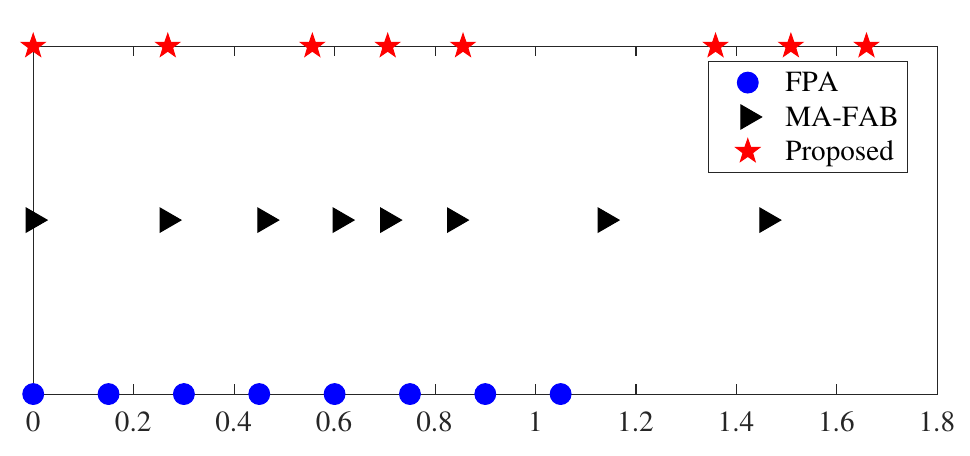}\label{apv3}}
\caption{(a) Optimized beam gain and (b) optimized MA positions for multi-region coverage.}
\vspace{-12pt}
\end{figure} 
	
Finally, by setting $K=1$ and $\theta_{\min}^{(1)}=0$, we plot the max-min beam gain versus $\theta_{\max}^{(1)}$ in Fig. \ref{max-min}. The number of MAs is set to $N = 8 $. It is observed that the max-min beam gains by all considered schemes monotonically decrease with ${\theta}_{\max}^{(1)}$. This is expected as the total beam gain over the entire spatial domain is a constant. Thus, when ${\theta}_{\max}^{(1)}$ or the coverage width increases, the max-min beam gain should reduce accordingly. It is also observed that in the case with a narrow-to-moderate coverage width, i.e., $\theta_{\max}^{(1)}\leq50^{\circ}$, the performance gap between the proposed scheme and the MA-FAB benchmark is negligible, while both of them significantly outperform the FPA benchmark. This indicates that adjusting the MA positions suffices to achieve uniform coverage for a narrow region given the proposed initial analog beamforming design. However, with the increase in ${\theta}_{\max}^{(1)}$, the performance gap between the proposed scheme and the MA-FAB benchmark becomes larger. This indicates that the analog beamforming design will play a more significant role for the more challenging wide-region coverage.

\begin{figure}[hbtp]
		\centering
		\includegraphics[scale=0.41]{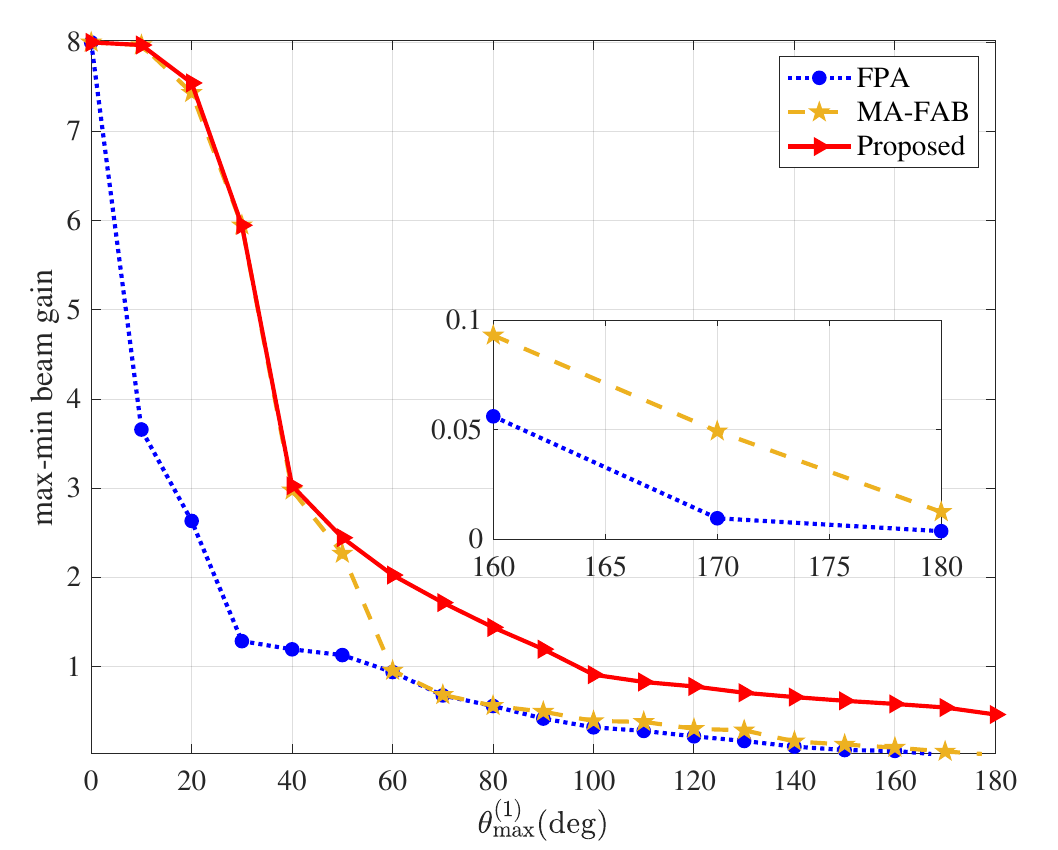}
		\caption{Max-min beam gain versus coverage width with $K=1$.}	
		\vspace{-9pt}
		\label{max-min}
\end{figure}

\section{Conclusion}
In this paper, we investigate the beam coverage design with multiple MAs by jointly optimizing their weights and positions to maximize the minimum beam gain over the desired spatial regions. An AO algorithm was proposed to obtain a high-quality suboptimal solution to this problem. Numerical results show that our proposed MA-enabled beam coverage design significantly outperforms the conventional FPAs by more flexibly catering to the geometry of the subregions. It was also shown that the antenna weights may play a more significant role as the width of the region increases.

\vspace{12pt}

\end{document}